\documentclass[pra,twocolumn,showpacs]{revtex4}

\usepackage{amssymb, amsmath}
\usepackage[dvips]{graphicx}
\usepackage{hyperref}

\newcommand{\ket}[1]{| #1 \rangle}

\newcommand{\oper}[2]{| #1 \rangle \langle #2 |}

\DeclareMathOperator{\Tr}{Tr}

\newtheorem{theorem}{Theorem}
\newtheorem{lemma}{Lemma}

\begin{document}
\title{Multipartite reduction criteria for separability}
\author{William Hall}
\email{wah500@york.ac.uk}
\affiliation{Department of Mathematics, University of York, Heslington, York YO10 5DD, UK}

\begin{abstract}
The reduction criterion is a well known necessary condition for separable states, and states violating this condition are entangled and also 1-distillable. In this paper we introduce a new set of necessary conditions for separability of multipartite states, obtained from a set of positive but not completely positive maps. These conditions can be thought of as generalisations of the reduction criterion to multipartite systems. We use tripartite Werner states as an example to investigate the entanglement detecting powers of some of these new conditions, and we also look at what these conditions mean in terms of distillation. Finally, we show that these maps can be used to give a partial solution to the \emph{subsystem problem}, as described in \cite{Sub}.
\end{abstract}


\maketitle
\section{Introduction}
Entanglement \cite{Ent} is one of the most intriguing phenomena in quantum mechanics. Its inherently non-classical nature caused much controversy in the early years of its discovery \cite{EPR, Bell}, and in more recent years many practical uses have been found for entanglement in the developing subject of quantum information science \cite{EntQI,NC}. This has led to a concerted effort to understand the nature of entanglement, and to create a consistent theory that allows us to both determine when an arbitrary mixed state is entangled, and to quantify the entanglement of that state. While the nature of bipartite pure state entanglement is well understood \cite{Ent1}, there is still much work to be done in the case of arbitrary mixed multipartite states; indeed, we still do not have an operational criteria to determine whether an arbitrary multipartite mixed state is entangled. However, in \cite{EntNC}, a necessary and sufficient condition for a bipartite state to be entangled was established using positive maps, and this was generalised to the multipartite case in \cite{EntNC_M}. We will make use of the positive map formalism heavily in this paper. A good review can be found in \cite{EntRev}.

The \emph{reduction criterion} \cite{Red, Red2} gives a necessary condition for a state to be separable; states which violate this criterion are hence entangled, and it can also be shown that these states are 1-distillable. In this paper we will produce some further necessary criteria for states to be (semi-)separable by introducing a new set of positive but not completely positive maps, and furthermore we will use the tripartite Werner states discussed in \cite{W3} to investigate their entanglement detecting properties. Finally, we will also discuss what the violation of these criteria means in terms of distillation. 

The criteria described in this paper are related to the subsystem compatibility problem: \emph{Given states of all proper subsystems of a multipartite system, what are the necessary and sufficient conditions for these subsystem states to be compatible with a single state of the whole system?} In \cite{Sub}, this problem is solved for a classical system of $n$ bits, and some of these conditions can be translated into necessary conditions for a system of $n$ qubits. We will show in this paper how our new set of positive maps can be used to derive some of these conditions in a more general setting.

\section{The reduction and generalised reduction criteria}
Let $\mathcal{H}$ be a Hilbert space, and $\mathcal{B}(\mathcal{H})$ be the set of (bounded) operators on $\mathcal{H}$. The reduction map $\Lambda: \mathcal{B}(\mathcal{H}) \to \mathcal{B}(\mathcal{H})$, which is defined by 
\begin{equation} \Lambda(\rho) = \Tr(\rho) \openone - \rho \end{equation}
($\openone$ is the identity matrix) is a positive map: If $\rho$ has positive eigenvalues $\lambda_i$ $(i = 1,\ldots,n)$ then $\Tr(\rho)= \sum_i \lambda_i$, and hence the eigenvalues of $\Lambda(\rho)$ are $\sum_{i \neq j} \lambda_i$ $(j = 1,\ldots,n)$, which are also positive. The map is not however completely positive: for instance, let $\ket{\Psi^+} = \frac{1}{\sqrt{2}}(\ket{00} + \ket{11}) \in \mathbb{C}^2 \otimes \mathbb{C}^2$, then 
\[  I \otimes \Lambda (\oper{\Psi^+}{\Psi^+}) = \frac{\openone}{2} - \oper{\Psi^+}{\Psi^+} \]
(where $I : \mathcal{B}(\mathcal{H}) \to \mathcal{B}(\mathcal{H})$ above is the identity map i.e. $I(\rho) = \rho)$ clearly has a negative eigenvalue. Hence for a bipartite state $\rho \in \mathcal{B}(\mathcal{H}_A \otimes \mathcal{H}_B)$,
\begin{equation} I \otimes \Lambda (\rho) = \rho_A \otimes \openone - \rho \geq 0 \end{equation}
(where $\rho_A = \Tr_B \rho$) is a necessary condition for a quantum state to be separable. This is known as the \emph{reduction criterion}. Any state violating this condition is not only entangled, but in fact is 1-distillable. This was originally proved in \cite{Red} and shown in a more abstract fashion in \cite{DW}.

Let us define the maps $\Lambda^{(n)} : \mathcal{B}(\mathcal{H}_1 \otimes \ldots \otimes \mathcal{H}_n) \to \mathcal{B}(\mathcal{H}_1 \otimes \ldots \otimes \mathcal{H}_n)$ by 
\begin{equation} \Lambda^{(n)}(\rho) = \sum_{B \subseteq N} (-1)^{|B|} \rho_B \label{GRM} \end{equation}
where $N=\{1,\ldots,n\}$, $|B|$ denotes the number of elements in the set $B$ and $\rho_B$ is the reduced density matrix of $\rho$ over the subsystems given in the set $B$, padded out with identities in the other systems (e.g. if $N=\{1,2,3\}$, and $B=\{1\}$, then $\rho_B = \Tr_{2,3}(\rho) \otimes \openone_2 \otimes \openone_3$). For example, 
\begin{eqnarray*}
\Lambda^{(1)}(\rho) &=& \Tr(\rho)\openone - \rho ;\\
\Lambda^{(2)}(\rho) &=& \Tr(\rho)\openone - \rho_1 - \rho_2 + \rho ;\\
\Lambda^{(3)}(\rho) &=& \Tr(\rho)\openone - \rho_1 - \rho_2 - \rho_3 + \rho_{12} + \rho_{13} + \rho_{23} - \rho.
\end{eqnarray*}
It is from these maps that we will form our necessary criteria for separability. We first note a preliminary result:

\begin{lemma} $\Lambda^{(n)}(\rho_1 \otimes \ldots \otimes \rho_n) = \bigotimes_{i=1}^n (\Tr(\rho_i)\openone - \rho_i)$ \end{lemma}

\emph{Proof} By induction: Trivially true for $n=1$; furthermore, assuming the result is true for the $n-1$ case,
\[\Lambda^{(n)}(\rho_1 \otimes \ldots \otimes \rho_n) = \sum_{B \subseteq N} (-1)^{|B|} \rho_B\]
\[= \sum_{B \subseteq N, 1 \notin B} (-1)^{|B|} \rho_B + \sum_{B \subseteq N, 1 \in B} (-1)^{|B|} \rho_B \]
\[= (\Tr(\rho_1)\openone - \rho_1) \otimes \sum_{B \subseteq N / \{1\}} (-1)^{|B|} \rho_B \]
\[= \bigotimes_{i=1}^n (\Tr(\rho_i)\openone - \rho_i) \ \Box \]

This result immediately shows the sense in which we can think about these maps as generalising the reduction criterion. We now prove that these maps are positive, and furthermore they satisfy a further condition known as \emph{$2n$-decomposability}. We first recall the definition of Schmidt number for density matrices \cite{Schmidt}: A bipartite density matrix $\rho$ has Schmidt number $k$ if 
\begin{enumerate}
\item For any decomposition $\rho = \sum_i p_i \oper{\psi_i}{\psi_i}$, at least one of the $\ket{\psi_i}$ has Schmidt rank (number of non-zero coefficients in Schmidt decomposition) at least $k$;
\item There exists a decomposition of $\rho$ with all vectors $\ket{\psi_i}$ of Schmidt rank no more than $k$.
\end{enumerate}
A map $\Lambda: \mathcal{B}(\mathcal{H}) \to \mathcal{B}(\mathcal{H})$ is $k$-decomposable if we can write $\Lambda = \Lambda_{CP} \circ T$, where $T$ is the transpose and $\Lambda_{CP}(\rho) = \sum_i V_i \rho V_i^\dagger$ is a completely positive map such that each $V_i$ has rank $k$.

With these definitions, we are ready to state our main result:

\begin{theorem} $\Lambda^{(n)}$ is a positive map, and is $2n$-decomposable. \end{theorem}

\emph{Proof} We make use of the Jamio\l kowski correspondence \cite{J}: Let $\mathcal{K} = \mathcal{H}_1 \otimes \ldots \otimes \mathcal{H}_n$ and consider the operator $A^{(n)} \in \mathcal{B}(\mathcal{K} \otimes \mathcal{K})$ defined by
\begin{equation} A^{(n)} = \left( I \otimes \Lambda^{(n)} \right) (P_+) \end{equation}
where
\begin{equation} P_+ = \sum_{i_1, \ldots, i_n, j_1, \ldots j_n} \oper{i_1 \ldots i_n i_1 \ldots i_n}{j_1 \ldots j_n j_1 \ldots j_n}, \end{equation}
a multiple of the projector onto the maximally entangled state on $\mathcal{K} \otimes \mathcal{K}$, and $\ket{i_r}$ forms an orthonormal basis for $\mathcal{H}_r$. We number the $2n$ systems as $\mathcal{H}_1, \ldots, \mathcal{H}_{2n}$ (so that $\mathcal{H}_{n+r} \cong \mathcal{H}_r$). Then, using Lemma 1,
\begin{eqnarray*}
A^{(n)} &=& \sum \oper{i_1 \ldots i_n}{j_1 \ldots j_n} \otimes \Lambda^{(n)} (\oper{i_1 \ldots i_n}{j_1 \ldots j_n} )\\
&=& \sum \oper{i_1 \ldots i_n}{j_1 \ldots j_n} \otimes \left( \bigotimes_{k=1}^n (\delta_{i_k j_k} \openone - \oper{i_k}{j_k})\right) \\
\end{eqnarray*}
Let us now write the above as an n-fold tensor product, grouping together the Hilbert spaces $\mathcal{H}_i$ for each $i$ (i.e. systems 1 and $n+1$ etc.). We can then write
\begin{eqnarray*}
A^{(n)} &=& \sum_{i_1, j_1} \left( \oper{i_1 j_1}{i_1 j_1} - \oper{i_1 i_1}{j_1 j_1} \right)_{1,n+1} \otimes \\
&& \ldots \otimes \sum_{i_n, j_n} \left(  \oper{i_n j_n}{i_n j_n} - \oper{i_n i_n}{j_n j_n} \right)_{n-1,2n} \\
&=&  \bigotimes_{k=1}^n \left( \openone - {P_+} \right)_{k,n+k} \\
\end{eqnarray*}
Let us consider the partial transpose of this over the second system $\mathcal{K}$ (i.e. over systems $n+1, \ldots, 2n$). Since
\begin{eqnarray*}
(\openone - {P_+})_{k,n+k}^{T_{n+k}} &=& \openone_{k,n+k} - \sum_{i_k, j_k} \oper{i_k j_k}{j_k i_k}_{k,n+k} \\
&=& \openone_{k,n+k} - {V}_{k,n+k}
\end{eqnarray*}
where $V$ is the swap operator, we have
\begin{equation}(A^{(n)})^{T_B} = \bigotimes_{k=1}^n \left( \openone - {V} \right)_{k,n+k} \label{witness} \end{equation}
where $T_B$ indicates partial transpose over the systems $n+1$ to $2n$. This is a positive operator, as $V$ has eigenvalues $\pm 1$.  Furthermore, $\openone - V$ has a Schmidt rank 2 decomposition: defining $\ket{\psi_{ij}} = \ket{ij} - \ket{ji}$, then
\begin{eqnarray*}
\sum_{i<j} \oper{\psi_{ij}}{\psi_{ij}} &=& \sum_{i \neq j} \oper{ij}{ij} - \oper{ij}{ji} \\
&=& \openone - V .
\end{eqnarray*}
Hence $(\openone-V)_{k,n+k}$ is a Schmidt rank 2 operator (the operator is clearly not separable and hence cannot have a Schmidt rank 1 decomposition). Due to the the tensor product structure of $A^{(n)}$ in (\ref{witness}), it must have Schmidt rank $2n$. 

It is possible to obtain $\Lambda^{(n)}$ from $A^{(n)}$ using the inversion formula \cite{J,DW}
\begin{equation} D = \left( I \otimes \Lambda \right)(P_+) \Leftrightarrow \Lambda(\rho) = \Tr_A \left[ D (\rho^T \otimes \openone) \right]. \end{equation}
Suppose $D = \left( \oper{\psi}{\psi} \right)^{T_B}$, with $\ket{\psi}$ having Schmidt decomposition $\sum_{i=1}^n c_i \ket{a_i, b_i}$ i.e. Schmidt rank is $n$. Some elementary algebra and the above inversion formula allows us to obtain that $\Lambda(\rho) = V \rho^T V^\dagger$, with $V=\sum_{i=1}^n c_i \oper{b_i^*}{a_i}$, an operator of rank $n$, where if $\ket{b_i} = \sum_m b_{im} \ket{m}$, then $\ket{b_i^*} = \sum_m b_{im}^* \ket{m}$. Hence by linearity, for $D^{T_B}$ being positive and having Schmidt rank $k$, $\Lambda$ is $k$-decomposable. It follows that $\Lambda^{(n)}$ is a $2n$-decomposable map. $\Box$

From these conditions we can obtain some of the necessary conditions mentioned briefly in the introduction to this paper. For odd $n$, the coefficient of $\rho$ in the expression for $\Lambda^{(n)}(\rho)$ is negative, and hence the expression
\begin{equation} \Lambda^{(n)}(\rho) + \rho = \sum_{B \subset N} (-1)^{|B|} \rho_B \end{equation}
is positive. Furthermore, this expression is written only in terms of the reduced density matrices of all of the proper subsystems of an $n$-partite system. Hence if the above expression is \emph{not} positive, it follows that the given reduced density matrices are not compatible with an overall state $\rho$. This generalises part of the necessary condition in \cite{Sub}, which is made for $n$-partite systems of qubits with $n$ odd, to the case of $n$-partite systems (still for odd $n$) where each individual system has arbitrary (finite) dimension.

\section{Entanglement detection}
We note that none of these maps are completely positive; this is evident from the construction above. An example \footnote{This was discovered in the $n=3$ case by Paul Butterley.} is $\mathcal{H}_i = \mathcal{H} \equiv \mathbb{C}^2$ for $i = 1,\ldots, n+1$ (i.e. $n+1$ qubits), and $\rho \in \mathcal{B}(\mathcal{H}^{\otimes (n+1)})$ defined by $\rho=\oper{\psi}{\psi}$, with $\ket{\psi} = \frac{1}{\sqrt{2}} (\ket{00}_{12} + \ket{11}_{12}) \otimes \ket{0}_{3\ldots n+1}$. Then, using (\ref{GRM}), 
\[I_1 \otimes \Lambda^{(n)}_{2\ldots n+1} (\rho) = \sum_{B \subset \{2,\ldots, n+1\}} (-1)^{|B|} \rho_{B \cup \{1\}} \]
\[= (\rho_1 \otimes \openone_2 - \rho_{12}) \otimes \sum_{B \subset \{3,\ldots, n+1\}} (-1)^{|B|} \rho_B \]
\[= (\rho_1 \otimes \openone_2 - \rho_{12}) \otimes \Lambda^{(n-1)}_{3\ldots n+1} (\oper{0}{0}_{3\ldots n+1}) \]
\[= (\openone_{12}/2 - \oper{\psi_+}{\psi_+}) \otimes \Lambda^{(n-1)}_{3\ldots n+1} (\oper{0}{0}_{3\ldots n+1}) \]
which has negative eigenvalues (the subscripts label which systems each map acts on).
We can hence think of each of these maps as providing necessary conditions for a density matrix to be separable: for a multipartite system with $n$ parts, and $A \subseteq N=\{1,\ldots,n\}$ with $|A|=k$, 
\begin{equation} I_{N \setminus A} \otimes \Lambda^{(k)}_A (\rho) \geq 0 \end{equation}
is a necessary condition for separability, where the subscript set denotes the systems that the map operates on. For example, for a tripartite state $\rho$, we have that
\begin{equation} \rho_1 - \rho_{12} - \rho_{13} + \rho \geq 0 \end{equation}
is a necessary condition for $\rho$ to have the semi-separable form $\rho= \sum_i p_i \rho_1^i \otimes \rho_{23}^i$. 

We note that if $I_1 \otimes \Lambda^{(n-1)}_{2\ldots n}(\rho) \ngeq 0$, then $I_1 \otimes \Lambda^{(n)}_{2\ldots n+1}(\rho \otimes \sigma) = I_1 \otimes \Lambda^{(n-1)}_{2\ldots n}(\rho) \otimes (Tr(\sigma)\openone -\sigma)$, which will also have negative eigenvalues.

One issue we are concerned with is the entanglement detecting power of these maps, especially in relation to the reduction criterion. The important result we will show here is that \emph{there are states detected by $\Lambda^{(2)}$ that are not detected by $\Lambda^{(1)}$, and vice versa}. To do this we utilise the tripartite Werner states that are introduced in \cite{W3} (we choose these states in particular as the entangled two-party Werner states are not detected by the reduction criterion \cite{Red}). 

Let us consider the Hilbert space $(\mathbb{C}^d)^{\otimes 3}$, and let us define the permutation operators
\[ V_\pi (\ket{\phi_1} \ket{\phi_2} \ket{\phi_3}) = \ket{\phi_{\pi^{-1}(1)}} \ket{\phi_{\pi^{-1}(2)}} \ket{\phi_{\pi^{-1}(3)}} \]
where $\pi \in S_3$. Then \cite[Lemma 1]{W3} states that the tripartite Werner states are given by $\rho = \sum_\pi \mu_\pi V_\pi (\mu_\pi \in \mathbb{C})$. We can rewrite these states using the following linear combinations, which we obtain from the representation theory of the group $S_3$:
\begin{eqnarray*}
R_+ &=& \frac{1}{6} \left( \openone + V_{(12)} + V_{(13)} + V_{(23)} + V_{(123)} + V_{(132)} \right)  \\
R_- &=& \frac{1}{6} \left( \openone - V_{(12)} - V_{(13)} - V_{(23)} + V_{(123)} + V_{(132)} \right)  \\
R_0 &=& \frac{1}{3} \left( 2\openone - V_{(123)} - V_{(132)} \right) \\
R_1 &=& \frac{1}{3} \left( 2V_{(23)} - V_{(13)} - V_{(12)} \right) \\
R_2 &=& \frac{1}{\sqrt{3}} \left( V_{(12)} - V_{(13)} \right) \\
R_3 &=& \frac{i}{\sqrt{3}} \left( V_{(123)} - V_{(132)} \right)
\end{eqnarray*}
We note that $R_+$ and $R_-$ represent the projections onto symmetric and antisymmetric subspaces (trivial and alternating representations of $S_3$), $R_0$ the projection onto the orthogonal subspace corresponding to the two-dimensional representation of $S_3$, and that $R_i \ (i = 1,2,3)$ act as the Pauli matrices in this subspace. This leads to Lemma 2 of \cite{W3}:

\begin{lemma} Let $\rho = \sum_k c_k R_k \ (k \in \{+,-,0,1,2,3\})$, and define $r_k(\rho) = tr(\rho R_k)$. Then $\rho$ is a density matrix if and only if
\[ r_+,r_-,r_0 \geq 0, \ r_+ + r_- + r_0 = 1, \ r_1^2 + r_2^2 + r_3^2 \leq r_0^2. \] \end{lemma}

We will consider a subset of these states parameterised by two real variables $a,b$. Let $\rho = \frac{1}{N}(a R_+ + (1-a)R_0 + bR_1)$. For positivity we require $0 \leq a \leq 1$, and $|r_1| \leq r_0$, which is equivalent to $|b|\leq a$. $N=\Tr(\rho)$ is the normalisation factor, and is given by 
\[ N = \frac{1}{6}d(d+1)(3a(2-d) + 4(d-1)). \]
We will consider the eigenvalues of  $I_{12} \otimes \Lambda^{(1)}_3(\rho) = \rho_{12} - \rho$ and $I_{2} \otimes \Lambda^{(1)}_{23}(\rho) = \rho_1 - \rho_{12} - \rho_{13} + \rho$. If $\rho$ is an entangled state, we are interested in the \emph{sign} of the eigenvalues: if all the eigenvalues are positive, then any present entanglement is not detected; one negative eigenvalue detects the entanglement. We note that $N$ is independent of $a$ if $d=2$, and for $d>2$, $N<0$ if and only if $a > \frac{4}{3} \left( 1+ \frac{1}{d-2} \right) > 1$, and hence we can ignore this normalisation factor for the purposes of determining the sign of the eigenvalues. Some elementary algebra outlined in Appendix A allows us to determine the eigenvalues of $\rho_{12} - \rho$ as
\begin{equation} \begin{array}{c}
\frac{1}{3N}(d-1)(2-a-b), \\
\frac{1}{3N}(d+1)(2-2a+b), \\
\frac{1}{6N}\bigg(a(d+2)+(1-a)(4d-6)+2b) \pm \\
 \Big\{(1/4)(a(d+2)-4(1-a)+b(12-2d))^2 +
\\ (3/4)(a(d+2)-4(1-a)-2bd)^2 \Big\}^{1/2} \bigg) 
\end{array} \label{e1} \end{equation}
and the eigenvalues of $\rho_1 - \rho_{12} - \rho_{13} + \rho$ as
\begin{equation} \begin{array}{c}
\frac{1}{6N} ( a((d+2)(d-3)+6) + 4(1-a)(d-1)^2 \\
+ 4b(d-1) ), \\
\frac{1}{6N} ( a(d+1)(d+2) - 4(1-a)(d+1)(d-3) \\
- 4b(d+1) ), \\
\frac{1}{6N} ( ad(d+2) + 2(1-a)(2(d+1)(d-1) - 4d +1) \\
 + 2b(1-d) ) ), \\
\frac{1}{6N} ( a(d+2)(d-2) + 2(1-a)(2(d+1)(d-1) - 4d + 5) \\
+ b(2d-10) ) 
\end{array} \label{e2} \end{equation}
(we note that for $d=2, R_-=0$ and hence the second expression in both lists above is \emph{not} an eigenvalue of the corresponding operator).
To determine which states are detected by which maps, we need only look at when the above eigenvalues are positive/negative and hence we can ignore the normalisation $1/N$, since for $0 \leq a \leq 1$, $N>0$. Below is a figure giving an example for $d=2$. It clearly shows states detected by $\Lambda^{(2)}$ but not $\Lambda^{(1)}$, and vice-versa. 

\begin{figure}[!ht] \begin{center} \includegraphics[scale=0.6]{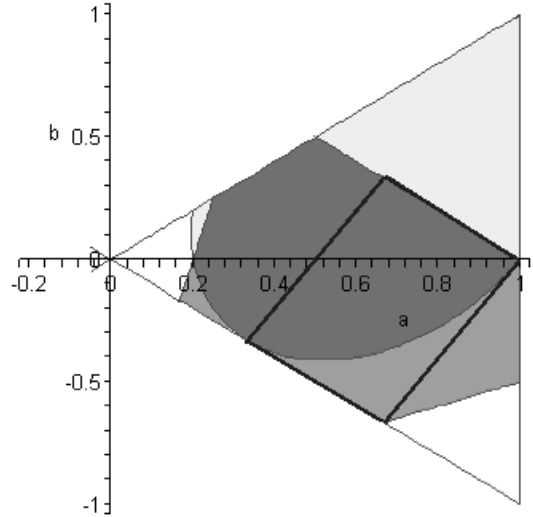} \end{center} \caption{\footnotesize Plot for $d=2$ illustrating which maps detect the state $\rho$ for given values of $a,b$. The white region represents states detected by both maps (i.e. have a negative eigenvalue), the lightest grey represents states detected by $\Lambda^{(2)}$ but not $\Lambda^{(1)}$; the middle grey those detected by $\Lambda^{(1)}$ but not $\Lambda^{(2)}$, and the darker grey is where neither map detects the state. The inside of the outlined box on the right of the graph represents states biseparable in the $1|23$ cut (this can be determined from \cite{W3}) and hence $I_{1} \otimes \Lambda^{(2)}_{23}(\rho)$ will never detect these states. } \end{figure}

Interestingly, for the above states for $d>2$ all states detected by $\Lambda^{(2)}$ are also detected by $\Lambda^{(1)}$. However no definite conclusions can be drawn from this since we are only looking at a small subset of symmetric states.

\section{Distillation criteria}
It is a well known result that if a bipartite state violates the reduction criterion then it is 1-distillable. This was first proved in \cite{Red} by giving an explicit distillation protocol and later in \cite{DW} using a more abstract positive map formalism. In this section we investigate the potential for the maps introduced above to be used as a distillation criterion. 

Consider a bipartite state $\rho \in \mathcal{B}(\mathcal{H}_A \otimes \mathcal{H}_B)$. We recall the following theorem from \cite{DW}:

\begin{theorem} $\rho$ is 1-undistillable if and only if $(I \otimes \Lambda)(\rho) \geq 0$ for all 2-decomposable maps $\Lambda: \mathcal{B}(\mathcal{H}_B) \to \mathcal{B}(\mathcal{H}_B)$. \end{theorem}

The reduction map $\Lambda^{(1)}$ is 2-decomposable, and so if $\rho$ satisfies the relation
\begin{equation} \rho_A \otimes \openone - \rho \ngeq 0 \label{re} \end{equation}
then $\rho$ is 1-distillable. This gives us the distillation criterion at the start of this article. Now, let us consider $\rho^{\otimes n} \in \mathcal{B}(\mathcal{H}_A^{\otimes n} \otimes \mathcal{H}_B^{\otimes n})$. Let us act $\Lambda^{(n)}$ on $\mathcal{H}_B^{\otimes n}$, and the identity map on $\mathcal{H}_A^{\otimes n}$. We wish to consider what the condition
\begin{equation} I_{1\ldots n} \otimes \Lambda^{(n)}_{n+1 \ldots 2n} (\rho^{\otimes n}) \ngeq 0 \label{newred} \end{equation}
means in terms of the distillation of $\rho$. We can deduce this from the following lemma:

\begin{lemma} $I_{1\ldots n} \otimes \Lambda^{(n)}_{n+1 \ldots 2n} (\rho^{\otimes n}) = (I \otimes \Lambda (\rho))^{\otimes n}$ \end{lemma}

\emph{Proof} Again by induction. Trivial for $n=1$; assuming the $n-1$ case, 
\[ I_{1\ldots n} \otimes \Lambda^{(n)}_{n+1 \ldots 2n} (\rho^{\otimes n})= \sum_{B \subset \{n+1, \ldots, 2n \} } (-1)^{|B|} \rho^{\otimes n}_{B \cup \{1, \ldots, n \} } \]
\[= ( {\Tr}_B( \rho ) \otimes \openone - \rho)_{1,n+1} \otimes \sum_{B \subset \{n+2, \ldots, 2n\}} (-1)^{|B|} \rho^{\otimes (n-1)}_{B \cup \{2, \ldots, n \}} \]
\[= (\Tr_B(\rho) \otimes \openone - \rho)_{1,n+1} \otimes \left( I_{2\ldots n} \otimes \Lambda^{(n)}_{n+2 \ldots 2n} (\rho^{\otimes (n-1)}) \right) \]
and hence the result holds by induction. $\Box$

From this result it is clear that
\begin{eqnarray*} 
I_{1\ldots n} \otimes \Lambda^{(n)}_{n+1 \ldots 2n} (\rho^{\otimes n}) \ngeq 0 & \Leftrightarrow & (I \otimes \Lambda (\rho))^{\otimes n} \ngeq 0 \\
&\Leftrightarrow& (I \otimes \Lambda (\rho)) \ngeq 0 
\end{eqnarray*}
and so condition (\ref{newred}) is equivalent to (\ref{re}), the violation of the reduction criterion.

\section{Conclusion}
In this paper we have given a set of positive but not completely positive maps that can be used to define a new set of separability criteria, and we have shown that they can be thought of as multi-party forms of the reduction criterion. We have analysed the entanglement detecting powers of these maps, and have shown that there are states that $\Lambda^{(2)}$ detects but not $\Lambda^{(1)}$, and vice-versa. It is highly likely that there will be similar results for the maps in general, and in theory, by using $n$-party analogues of the Werner states, we could obtain further results. It remains to be seen (although from the above results it seems unlikely) whether there is any hierarchy within these separability criteria. We have also analysed the criteria from a distillation viewpoint, and have shown that we can recover the criteria for distillation that we obtain from the reduction criterion.

\begin{acknowledgments}
The author would like to thank Paul Butterley for providing the impetus behind the above work, Matthias Christandl for a very illuminating conversation, Lieven Clarisse and Anthony Sudbery for checking the final manuscript, and finally the Engineering and Physical Sciences Research Council (UK) for supporting this work.
\end{acknowledgments}

\appendix
\section{Details of calculations for tripartite Werner states}
In this appendix we give a few more details of the calculations required to obtain the eigenvalues of the given maps in section 3.
Suppose that $\rho = \sum_k c_k R_k \ (k \in \{+,-,0,1,2,3\})$. Then defining
\[ \nu_+ = \frac{d}{6}(d^2 + 3d+2), \ \nu_- = \frac{d}{6}(d^2 - 3d+2), \ \nu_0 = \frac{d}{3}(d^2 - 1),\]
it can easily be shown that $r_+ = c_+ \nu_+, r_- = c_- \nu_-, r_i = 2 c_i \nu_0 \ (i \in \{0,1,2,3\})$. Furthermore, for $k=+,-,0$, the projectors $R_k$ are orthogonal, and project onto a subspace of dimension $\nu_k$, and $R_1,R_2,R_3$ act as Pauli matrices within the subspace of $R_0$. Hence the eigenvalues of $\rho$ are given by
\[ c_+ \textrm{ (multiplicity }\nu_+), \ c_- \textrm{ (multiplicity }\nu_-) \]
\begin{equation} c_0 \pm \sqrt{c_1^2 + c_2^2 + c_3^3} \textrm{ (multiplicity }\nu_0) \label{eig} \end{equation}
This allows us to easily investigate the properties of $I_{12} \otimes \Lambda^{(1)}_3(\rho)$ and $I_{2} \otimes \Lambda^{(1)}_{23}(\rho)$. 
Now take $\rho = \frac{1}{N}(a R_+ + (1-a)R_0 + bR_1)$
with $N = \frac{1}{6}d(d+1)(3a(2-d) + 4(d-1))$
 as defined above. We reiterate that for positivity we require $0 \leq a \leq 1$, and $|b|\leq a$.  
Inverting the relations for $R_+$ etc. in terms of $V_\rho$, and some tedious but elementary algebra allows us to write 
\begin{eqnarray*}
\lefteqn{\rho_{12} - \rho =} \\
 && \frac{1}{3N}(d-1)(2-a-b)R_+ \\
 &+& \frac{1}{3N}(d+1)(2-2a+b)R_- \\
 &+& \frac{1}{6N} (a(d+2) + (1-a)(4d-6) + 2b)R_0 \\
 &-& \frac{1}{12N} (a(d+2)-4(1-a)+b(12-2d))R_1 \\
 &+& \frac{\sqrt{3}}{12N}(a(d+2)-4(1-a)-2bd)R_2 
\end{eqnarray*}
and 
\begin{eqnarray*}
\lefteqn{\rho_1 - \rho_{12} - \rho_{13} + \rho =} \\
&& \frac{1}{6N}  a((d+2)(d-3)+6)  \\
&& + 4(1-a)(d-1)^2 + 4b(d-1) )R_+  \\
 &+& \frac{1}{6N} ( a(d+1)(d+2) - 4(1-a)(d+1)(d-3) \\
 && - 4b(d+1) )R_- + \frac{1}{6N} ( a(d-1)(d+2) \\
 && + (1-a)(4(d-1)(d+1) - 8d + 6) - 4b)R_0 \\
 &+& \frac{1}{6N} ( a(d+2) - 4(1-a) - b(6-2d) ) R_1 
\end{eqnarray*}
from which, using (\ref{eig}) above also, we can easily determine the eigenvalues of $\rho_{12} - \rho$ and $\rho_1 - \rho_{12} - \rho_{13} + \rho$ given in equations (\ref{e1}) and (\ref{e2}) above.
We finish by noting that we do not need to consider $\rho_{13} - \rho$ additionally, as it can be shown to have identical eigenvalues to $\rho_{12} - \rho$ (this follows simply from the symmetry of $\rho$ between systems 2 and 3, and the symmetry of the expressions for $V_\rho$ in terms of $R_+$ etc.).

\end{document}